\documentclass{ws-procs9x6}

\usepackage{epsfig}

\newcommand{\nn}{\nonumber}

\newcommand{\be}{\begin{equation}}
\newcommand{\ee}{\end{equation}}
\newcommand{\bea}{\begin{eqnarray}}
\newcommand{\eea}{\end{eqnarray}}

\begin{document}

\title{Using $\nu_{\rm e} \to \nu_\tau$: golden and silver channels
at the Neutrino Factory}

\author{A.~Donini\footnote{\uppercase{W}ork
supported by the \uppercase{P}rograma \uppercase{R}am\'on 
y \uppercase{C}aj\'al of
the \uppercase{M}inist\'erio de \uppercase{C}iencia y 
\uppercase{T}ecnolog\'{\i}a of \uppercase{S}pain.}}

\address{Instit\'uto F\'{\i}sica Te\'orica \\
C-XVI, Universidad Autonoma Madrid \\ 
Cantoblanco, E-28049, Madrid, Spain\\ 
E-mail: andrea.donini@roma1.infn.it}


\maketitle

\abstracts{
I briefly review the source of the so-called intrinsic ambiguity and show how the combination
of ``golden'' and ``silver'' channel at the Neutrino Factory can solve the problem,
in the absence of other sources of degeneracies. I then relaxed the hypothesis 
$\theta_{23} = 45^\circ$ and show how the different dependence of the two 
channels on $\theta_{23}$ can help in solving the intrinsic and $\theta_{23}$-octant
ambiguity at the same time. 
}



%


The most sensitive method to study leptonic CP violation \cite{Cervera:2000kp}
is the measure of the transition probability 
$\nu_e(\bar \nu_e) \rightarrow \nu_\mu(\bar \nu_\mu)$. 
In the framework of a Neutrino Factory-based beam \cite{Apollonio:2002en} this is called 
the ``{\it golden channel}'': being an energetic electron neutrino 
beam produced with no contamination from muon neutrinos with the same helicity 
(only muon neutrinos of opposite helicity are present in the beam), 
the transition of interest can be easily measured by searching for 
wrong-sign muons, i.e. muons with charge opposite to that of the muons in the
storage ring, provided the considered detector has a good muon charge 
identification capability. 
The transition probability $\nu_e \to \nu_\tau$ is also extremely sensitive \cite{Donini:2002rm}
to the leptonic CP-violating phase $\delta$. We can indeed look for muonic decay of wrong-sign $\tau$'s 
(the so-called ``{\it silver channel}'', due to its lesser statistical significance with respect
to the ``golden channel'') in combination with wrong-sign muons from $\nu_e \to \nu_\mu$ 
to improve our measurement. 

The transition probability (at second order in perturbation theory in
$\theta_{13}$, $\Delta_\odot/\Delta_{atm}$, $\Delta_\odot/A$ and
$\Delta_\odot L$) is \cite{Burguet-Castell:2001ez,Freund:2001pn,Minakata:2002qe}:
\be
\label{eq:spagnoli}
P^\pm_{e \mu} (\bar \theta_{13}, \bar \delta) = X_\pm \sin^2 (2 \bar
\theta_{13}) + Y_\pm \cos ( \bar \theta_{13} ) \sin (2 \bar
\theta_{13} ) \cos \left ( \pm \bar \delta - \frac{\Delta_{atm} L }{2}
\right ) + Z \, , \ee and \be
\label{eq:etau}
P^\pm_{e \tau} (\bar \theta_{13}, \bar \delta) = 
X^\tau_\pm \sin^2 (2 \bar \theta_{13}) -
Y_\pm \cos ( \bar \theta_{13} ) \sin (2 \bar \theta_{13} )
      \cos \left ( \pm \bar \delta - \frac{\Delta_{atm} L }{2} \right ) 
+ Z^\tau \, ,
\ee
where $\pm$ refers to neutrinos and antineutrinos, respectively.
The parameters $\bar \theta_{13}$ and $\bar \delta$ are the physical
parameters that must be reconstructed by fitting the experimental data
with the theoretical formula for oscillations in matter.  
The coefficients of the two equations are:
\bea
X_\pm &=& s^2_{23} 
\left ( \frac{\Delta_{atm} }{ B_\mp } \right )^2
\sin^2 \left ( \frac{ B_\mp L}{ 2 } \right ) \ , \qquad \qquad
X^\tau_\pm = \left ( c^2_{23}/s^2_{23} \right )  X_\pm \, , \nn \\
Y_\pm &=& \sin ( 2 \theta_{12} ) \sin ( 2 \theta_{23} )
\left ( \frac{\Delta_\odot }{ A } \right )
\left ( \frac{\Delta_{atm} }{ B_\mp } \right )
\sin \left ( \frac{A L }{ 2 } \right )
\sin \left ( \frac{ B_\mp L }{ 2 } \right ) \, , \\
Z &=& c^2_{23} \sin^2 (2 \theta_{12})
\left ( \frac{\Delta_\odot }{ A } \right )^2
\sin^2 \left ( \frac{A L }{ 2 } \right ) \ , \qquad 
Z^\tau = \left ( s^2_{23}/c^2_{23} \right ) Z \, , \nn 
\eea
with $A = \sqrt{2} G_F n_e$ (expressed in
eV$^2$/GeV), $B_\mp = | A \mp \Delta_{atm}|$ (with $\mp$ referring
to neutrinos and antineutrinos, respectively) and $\Delta_{atm}
= \Delta m^2_{23} / 2 E_\nu$, $\Delta_\odot = \Delta m^2_{12} / 2
E_\nu$. The parameters in $X, Y, Z$ have been considered as fixed
quantities, supposed to be known by the time when the Neutrino Factory
will be operational with good precision: we put $\theta_{12} = 33^\circ$ 
and $\Delta m^2_{12} = 1.0 \times 10^{-4} $ eV$^2$; $\theta_{23} = 45^\circ$ and 
$\Delta m^2_{23} = 2.9 \times 10^{-3} $ eV$^2$, with $\Delta m^2_{23}$ positive 
(for $\theta_{23} = 45^\circ$ the $\theta_{23}$-octant ambiguity \cite{Barger:2001yr} 
is absent); $A = 1.1 \times 10^{-4} $ eV$^2$/GeV (a good approximation $L < 4000$ Km).
We have not included errors on these parameters since the inclusion
of their uncertainties does not modify the results for $\theta_{13}$ 
and $\delta$ significantly \cite{Burguet-Castell:2001ez}.

Eqs.~(\ref{eq:spagnoli}) and (\ref{eq:etau}) lead to two
equiprobability curves in the ($\theta_{13}, \delta$) plane for
neutrinos and antineutrinos of a given energy: \be
\label{eq:equi0} 
P^\pm_{e \mu} (\bar \theta_{13}, \bar \delta) = P^\pm_{e \mu}
(\theta_{13}, \delta) \, ; \qquad \qquad P^\pm_{e \tau} (\bar \theta_{13},
\bar \delta) = P^\pm_{e \tau} (\theta_{13}, \delta) \, .  \ee 
Notice that $X^\tau_\pm$ and $Z^\tau$ differ from the corresponding
coefficients for the $\nu_e \to \nu_\mu$ transition for the $\cos
\theta_{23} \leftrightarrow \sin \theta_{23}$ exchange, only
(and thus for $\theta_{23} = 45^\circ$ we have $X = X^\tau, Z = Z^\tau$). 
The $Y_\pm$ term is identical for the two channels, but it appears with an
opposite sign. This sign difference in the $Y$-term is crucial, as it
determines a different shape in the $(\theta_{13}, \delta)$ plane for
the two sets of equiprobability curves.

\begin{figure}[h!]
\begin{center}
\hspace{-1cm} \epsfxsize7cm\epsffile{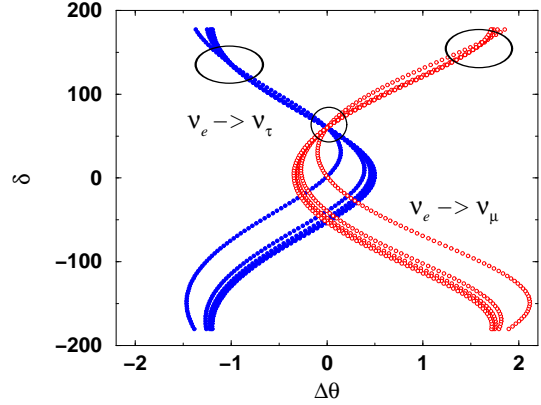} 
\caption{\it Equiprobability curves for neutrinos in the ($\Delta \theta, \delta$)
plane, for $\bar \theta_{13} = 5^\circ, \bar \delta = 60^\circ$,
$E_\nu \in [5, 50] $ GeV and $L = 732$ Km for the $\nu_e \to \nu_\mu$
and $\nu_e \to \nu_\tau$ oscillation. $\Delta \theta$ is the
difference between the input parameter $\bar \theta_{13}$ and the
reconstructed parameter $\theta_{13}$, $\Delta \theta = \theta_{13} -
\bar \theta_{13}$.}
\label{fig:equiprobetau}
\end{center}
\end{figure}

In Fig.~\ref{fig:equiprobetau} equiprobability curves for the $\nu_e \to \nu_\mu, \nu_\tau$ 
oscillations at a fixed distance, $L = 732$ Km, with input parameters
$\bar \theta_{13} = 5^\circ$ and $\bar \delta = 60^\circ$ and
different values of the energy, $E_\nu \in [5, 50]$ GeV, have been superimposed. 
The effect of the different sign in front of the $Y$-term in
eqs.~(\ref{eq:spagnoli}) and (\ref{eq:etau}) can be seen in the
opposite shape in the ($\theta_{13}, \delta$) plane of the $\nu_e \to
\nu_\tau$ curves with respect to the $\nu_e \to \nu_\mu$ ones.  Notice
that both families of curves meet in the ``physical'' point,
$\theta_{13} = \bar \theta_{13}$, $\delta = \bar \delta$, and any
given couple of curves belonging to the same family intersect in a
second point that lies in a restricted area of the ($\Delta \theta,
\delta$) plane, the specific location of this region depending on the
input parameters ($\bar \theta_{13}, \bar \delta$) and on the neutrino 
energy. Using a single set of experimental data (e.g. the ``golden'' muons), 
a $\chi^2$ analysis will therefore identify two allowed regions: the ``physical''
one (i.e. around the physical value, $\bar \theta_{13}, \bar \delta$) and
a ``clone'' solution, spanning all the area where a second
intersection between any two curves occurs.  This is the source of the
so-called intrinsic ambiguity \cite{Burguet-Castell:2001ez}. When
considering at the same time experimental data coming from both the
``golden'' and the ``silver'' channels, however, a comprehensive
$\chi^2$ analysis of the data would result in the low-$\chi^2$ region around the physical pair, 
only \cite{Donini:2002rm}, since ``clone'' regions for each set of data lie well apart. 

\begin{figure}[h!]
\begin{center}
\begin{tabular}{cc}
\hspace{-1cm} \epsfxsize5cm\epsffile{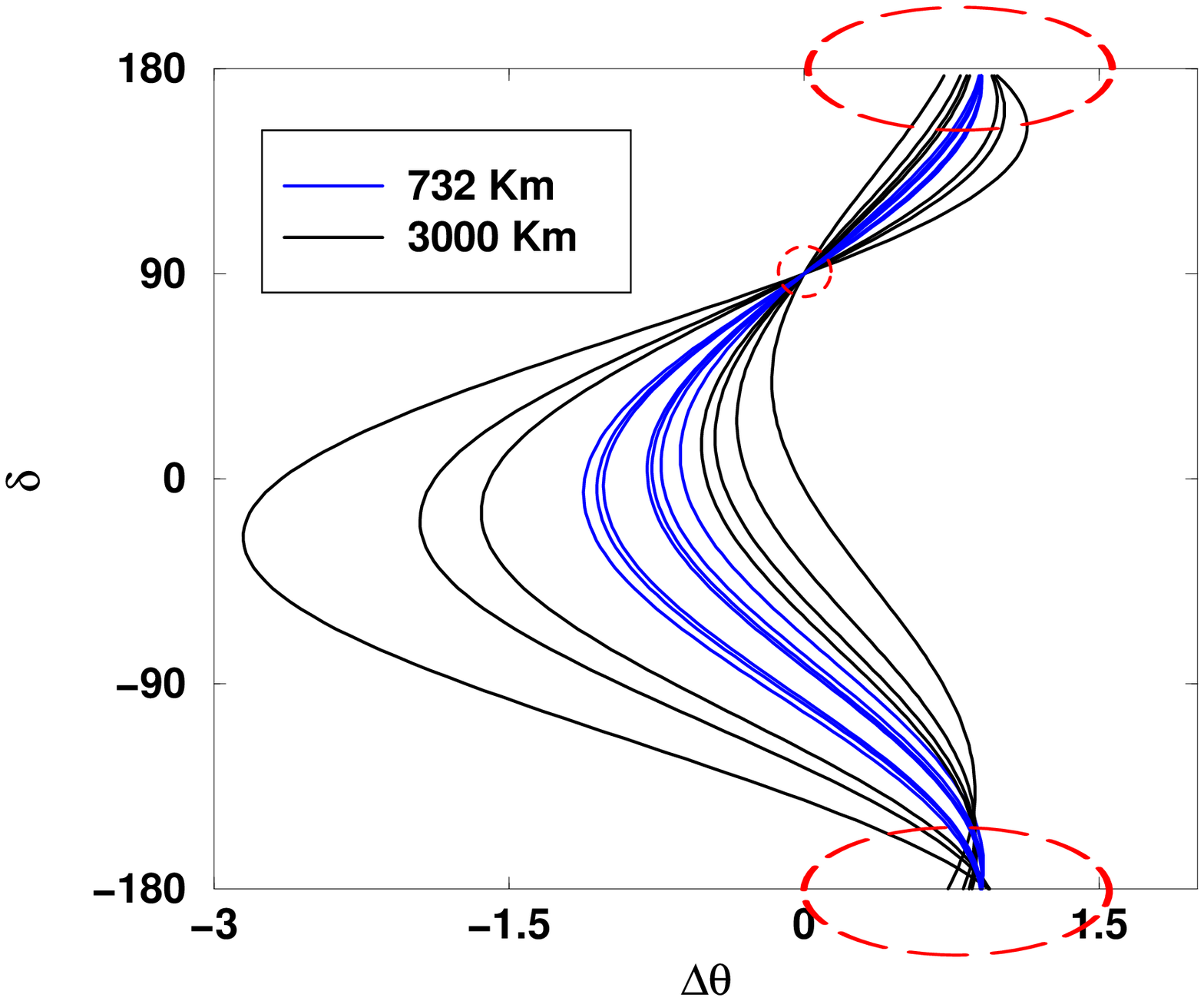} &  \epsfxsize4.5cm\epsffile{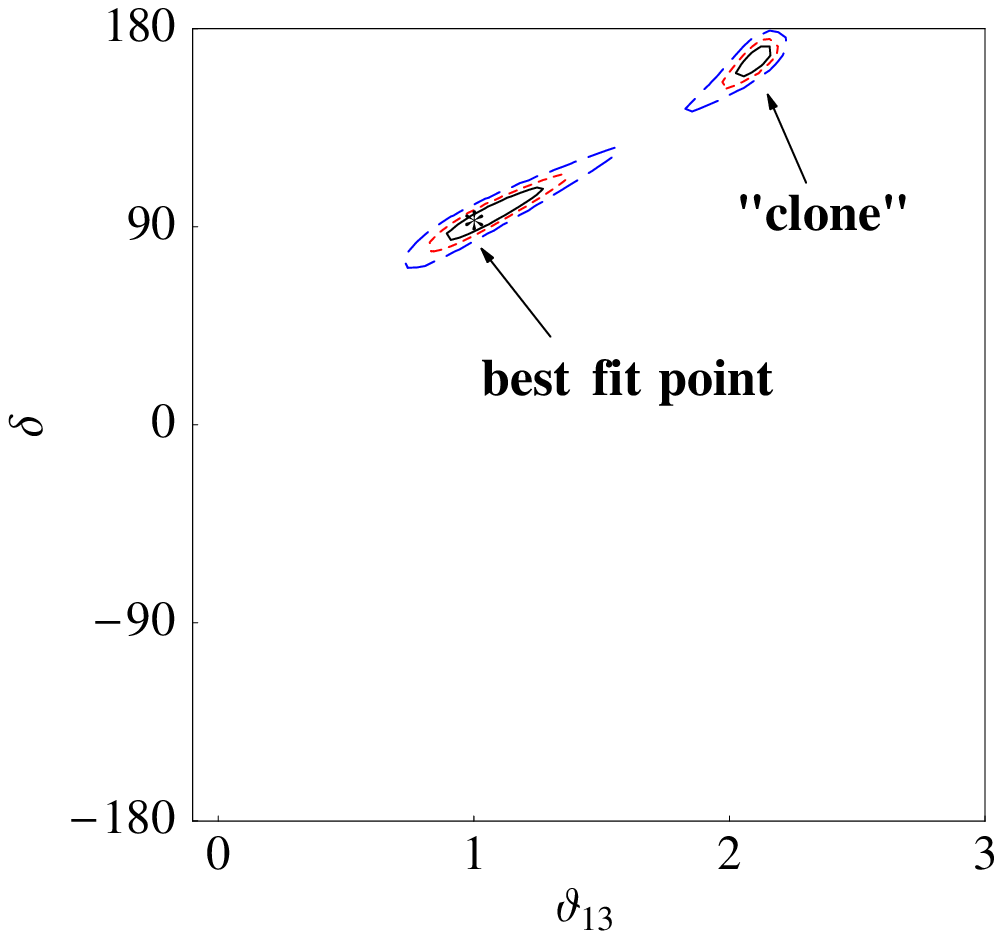} \\
\hspace{-1cm} \epsfxsize5cm\epsffile{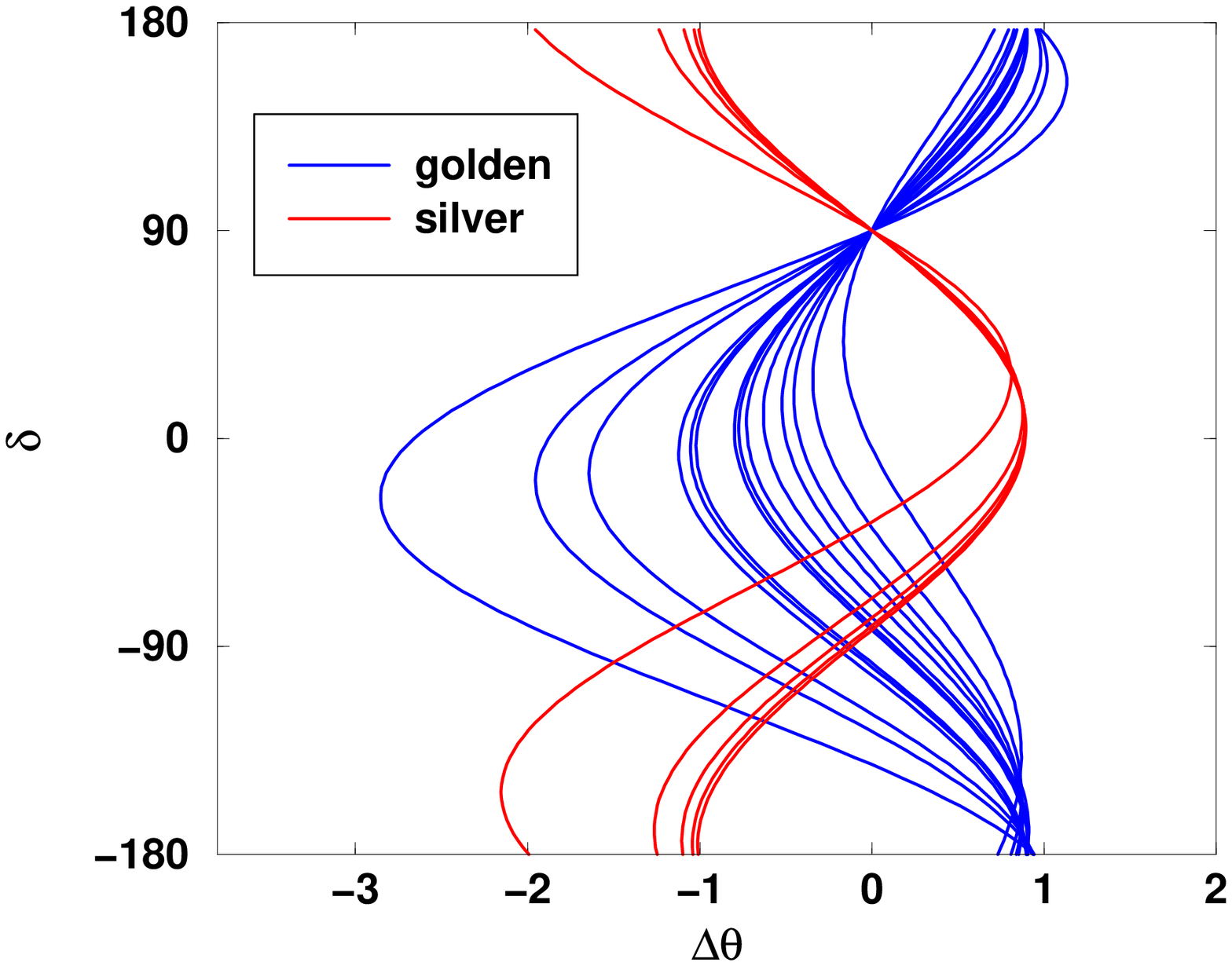} & \epsfxsize4.5cm\epsffile{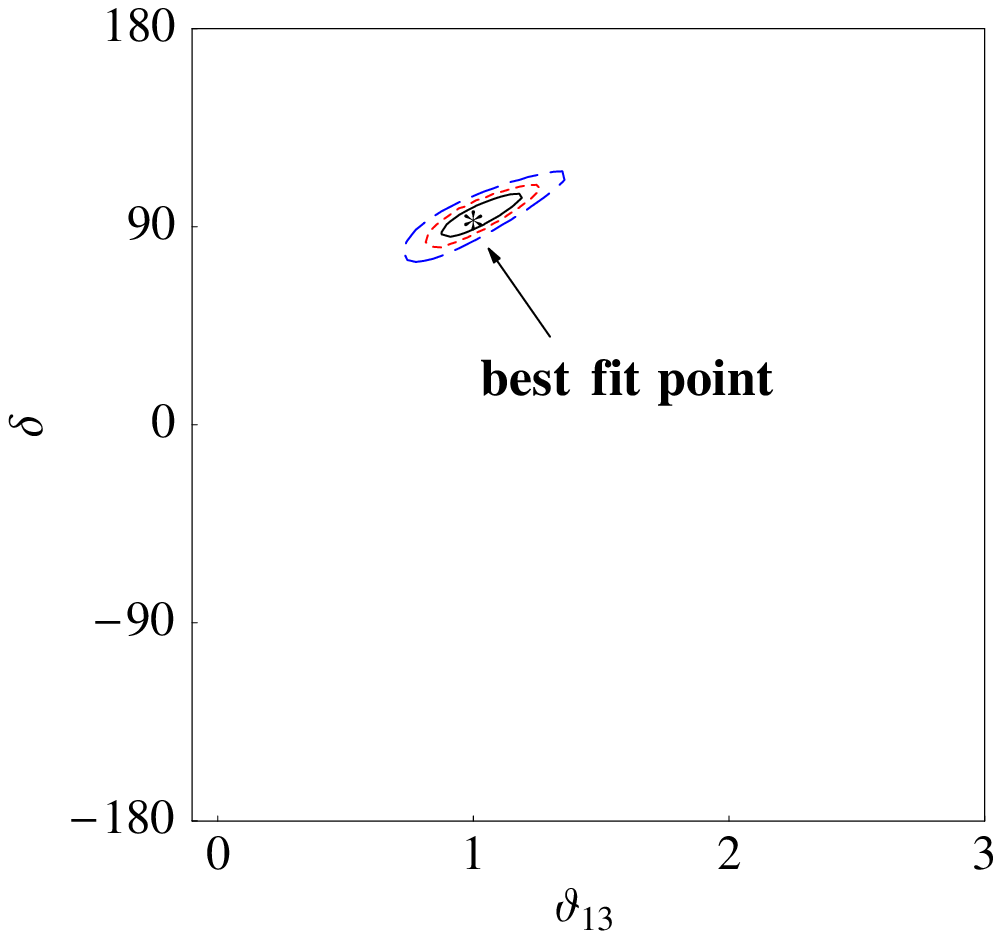}
\end{tabular}
\caption{\it 
Equiprobability curves for neutrinos and antineutrinos (left) and the corresponding outcome 
of a fit (right) for golden channel only (above) and including both the golden and the
silver channel (below). The input parameters are $\bar \theta_{13} = 1^\circ; \bar \delta = 90^\circ$.
The golden signal is obtained at a realistic 40 Kton MID
and at an ideal 2 Kton ECC. The results are substantially unchanged when
considering a realistic ECC of doubled size of 4 Kton.
}
\label{fig:ideal}
\end{center}
\end{figure}

This statement is only true if the statistical significance of both
sets of experimental data is sufficiently high. The golden channel 
has been thoroughly studied \cite{Cervera:2000kp}
considering a 40 Kton magnetized
iron detector \cite{Cervera:2000vy} (MID) located at $L = 3000$ Km.
A dedicated analysis of the silver channel at an OPERA-like \cite{proposal} 
Emulsion Cloud Chamber (ECC) detector has been recently performed \cite{Autiero:2003fu} 
to substantiate the results on the impact of the silver muons when combined
with golden muons \cite{Donini:2002rm} based on the OPERA proposal.
 
In Fig. \ref{fig:ideal} we present equiprobability curves and the outcome of the fit
when only golden muons (above) or both golden and silver muons (below) are considered, 
for input parameters $\bar \theta_{13} = 1^\circ; \bar \delta = 90^\circ$.
The golden muon signal is measured at a 40 Kton MID, with realistic efficiency 
and background \cite{Cervera:2000kp}; the silver muon signal is measured at an ideal 2 Kton ECC 
with spectrometers following the OPERA proposal \cite{proposal}. 
These results do not change when including efficiency and backgrounds if considering a 
doubled-size ECC of 4 Kton. Notice that
the scanning power needed to take full advantage of a 4 Kton detector is considered to be
easily under control by the time of the Neutrino Factory \cite{Autiero:2003fu}.

In order to study the effect of the $\theta_{23}$-octant
ambiguity, we now relax the hypothesis on the value of $\theta_{23} = 45^\circ$ (for which
no ambiguity was present). 

\begin{figure}[h!]
\begin{center}
\begin{tabular}{cc}
\hspace{-1cm} \epsfxsize6cm\epsffile{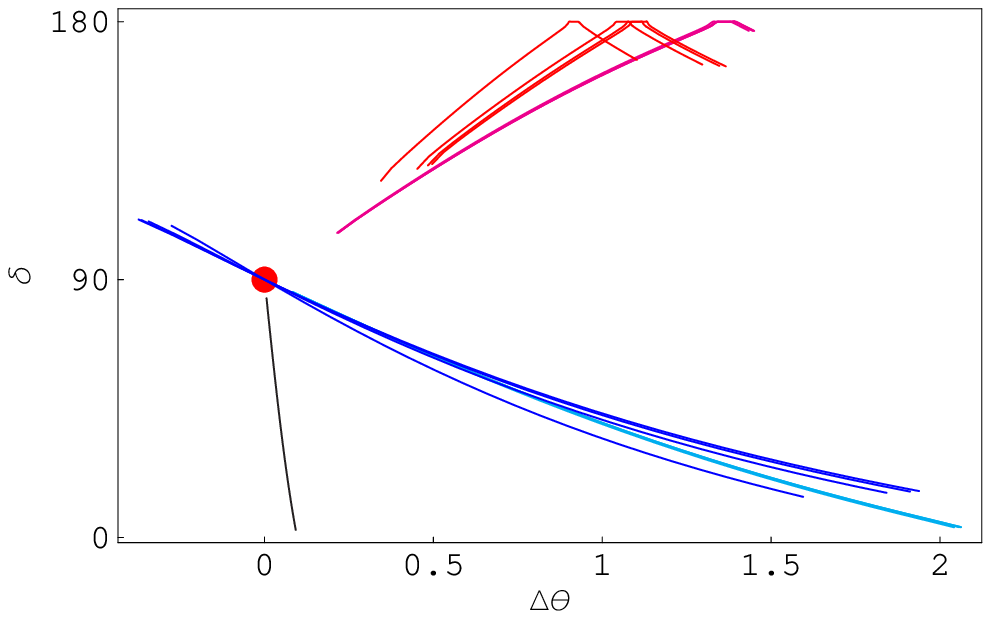} & 
              \epsfxsize6cm\epsffile{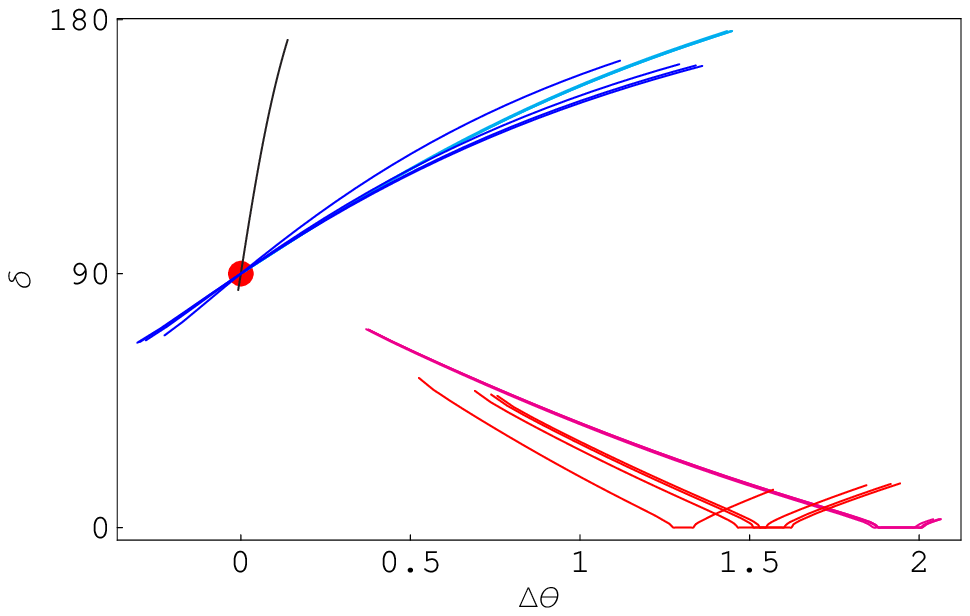}
\end{tabular}
\caption{\it The trajectory in the ($\Delta \theta, \delta$)
plane of the clone regions for $\bar \delta = 90^\circ$ as a function 
of $\bar \theta_{13}$ for the SPL at $L = 130$ Km and the Neutrino Factory 
at $L = 732, 3000$ Km. 
In the case of the Neutrino Factory, both the golden and the silver channel
are considered. The thick dot is the true solution, located at
$\Delta \theta = 0^\circ, \delta = \bar \delta = 90^\circ$.
The two plots represent: $\theta_{23} = 40^\circ$ (left); $\theta_{23} = 50^\circ$ (right).
}
\label{fig:traj90}
\end{center}
\end{figure}

It is possible to compute the analytical location of the clones by solving eq.~(\ref{eq:equi0}) 
for $\theta_{13}$ and $\delta$ as a function of the input parameters $\bar \theta_{13}, \bar \delta$,
after convoluting over the flux and the neutrino-nucleon cross-section and integrating over
the neutrino energy. 
In Fig. \ref{fig:traj90} we present the trajectory of the intrinsic clone region for 
$\bar \delta = 90^\circ$ for the different channels, beams and baselines in the 
($\Delta \theta, \delta$) plane as a function of $\bar \theta_{13} \in [0.01^\circ, 10^\circ]$ 
for the two cases of $\theta_{23} = 40^\circ$ (left) and $\theta_{23} = 50^\circ$ (right). 
The thick dot is the true solution, $\Delta \theta = 0^\circ, \delta = \bar \delta$.
In Fig. \ref{fig:traj90} (left, $\theta_{23} = 40^\circ$), the isolated line going down 
represents the displacement of the intrinsic clone for the SPL beam \cite{Mezzetto:2003mm} for the 
CERN-Frejus baseline, $L = 130$ Km.
The cluster of lines moving towards larger values of $\delta$ and $\Delta \theta$ represent
the displacement of the intrinsic clone for the golden channel at the Neutrino Factory beam, 
for 4 different energy bins (in the neutrino energy range $E_\nu \in [10, 50]$)
and two baselines, $L = 732$ and $L = 3000$ Km. Finally, the cluster of lines moving towards 
smaller values of $\delta$ and larger values of $\Delta \theta$ represent
the displacement of the intrinsic clone for the silver channel at the Neutrino Factory beam, 
again for 4 different energy bins and two baselines. Notice that the qualitative behaviour of 
golden and silver trajectories is substantially independent of the neutrino energy and of the 
considered baseline. 
In Fig. \ref{fig:traj90} (right, $\theta_{23} = 50^\circ$) the SPL trajectory moves up and 
the golden and silver cluster are interchanged with respect to the previous case, due to 
the interchange of $\cos \theta_{23} \leftrightarrow \sin \theta_{23}$ in the $X$ and $Z$
coefficients of eqs. (\ref{eq:spagnoli}) and (\ref{eq:etau}).

For decreasing values of $\bar \theta_{13}$, the clones move away from the thick dot, the physical
point. Notice that for large $\bar \theta_{13}$, all lines are relatively near the physical input pair:
this reflects the fact that for $\bar \theta_{13}$ large enough degeneracies are not a 
problem \cite{Huber:2002rs}, being $\Delta \theta / \theta_{13}$ always small.
For small values of $\bar \theta_{13}$, any combination of experiments (golden and silver at the 
Neutrino Factory, or any of the two in combination with the SPL superbeam \cite{Burguet-Castell:2002qx}) 
would result in killing
the intrinsic degeneracy, provided that statistics of the considered signals is large enough.
Regarding the $\theta_{23}$-octant ambiguity it seems reasonable to expect
that the combination of the golden and silver channels would help in solving the ambiguity,
due to the different behaviour of the respective clones depending on the value of $\theta_{23}$. 
This seems indeed to be the case from a preliminar analysis of our data \cite{Mena:2003yh} for 
$\bar \theta_{13} \geq 2^\circ$, with a loss of sensitivity below this value due to 
the extremely poor statistics in the silver channel. 

%
%


\end{document}